\documentclass[12pt,epsf]{article}

\usepackage{graphicx}
\usepackage{color}

\def \mmev   {meV/$c^{2}$}

\def \tchi1  {$\chi_{1}$} 
\def \tchi2  {$\chi_{2}$}

\renewcommand{\thebibliography}[1]{\subsubsection*{References}\list
  {[\arabic{enumi}]}{\settowidth\labelwidth{[#1]}\leftmargin\labelwidth
    \advance\leftmargin\labelsep\usecounter{enumi}}}

\begin{document}


\begin{center}
\begin{Large}
\bf{Search for Radiative Decays of Cosmic Background Neutrino  using Cosmic Infrared Background Energy Spectrum \\
}
\end{Large}
\end{center}

Shin-Hong Kim$^1$
, Ken-ichi Takemasa$^1$, Yuji Takeuchi$^1$, and Shuji Matsuura$^2$\\

\begin{small}

$^1$ Graduate school of Pure and Applied Sciences, University of Tsukuba,Tsukuba, Ibaraki, 305-8571, Japan\\
$^2$ Institute of Space and Astronautical Science, JAXA, Sagamihara, Kanagawa, 252-5210, Japan \\ 
\end{small}

\begin{center}
Abstract\\
\end{center}

       We propose to search for the neutrino radiative decay by fitting 
a photon energy spectrum of
the cosmic infrared
 background 
to a sum of the photon
 energy spectrum from the neutrino radiative decay and a continuum. 
By comparing
 the present cosmic infrared background energy spectrum observed by AKARI 
and Spitzer to the photon energy spectrum expected from neutrino 
radiative decay 
with a maximum likelihood method, we obatined 
a lifetime lower limit of $3.1 \times 10^{12}$  to 
$3.8 \times 10^{12}$ years
at 95\% confidence level
for the third generation neutrino $\nu_3$  
in the $\nu_3$  mass range between 50 \mmev\ and 150 \mmev\ 
under the present constraints by the neutrino oscillation measurements. 
In the left-right symmetric model, the minimum lifetime of $\nu_3$ is
predicted to be $1.5 \times 10^{17}$ years  
for $m_3$ of 50 \mmev.  We studied the feasibility of  the observation of the 
neutrino radiative decay 
 with a lifetime of $1.5 \times 10^{17}$ years, by measuring 
 a continuous energy spectrum of the cosmic infrared 
background.


\newpage
\section{Introduction}

 The difference between the mass-squares of different-generation neutrinos has been measured by 
atmospheric neutrino oscillation 
experiments with Super-Kamiokande \cite{SK1,SK2},  neutrino beam experiments with 
K2K \cite{K2K} and MINOS \cite{MINOS1,MINOS2}, 
solar neutrino 
oscillation experiments with Super-Kamiokande \cite{solar-SK} and 
SNO \cite{SNO1,SNO2} and a nuclear reactor oscillation 
experiment with KamLAND \cite{KamLAND1,KamLAND2}, but the neutrino mass itself has not 
been measured yet.  Detection of 
neutrino radiative decay enables us to measure a quantity independent of the difference 
between the mass-squares of different-generation neutrinos. Thus we can determine the neutrino 
mass itself from these two independent measurements, the neutrino oscillation and the neutrino 
radiative decay.

As the neutrino lifetime is so long as to be much larger than the age of the universe, 
the most promising method is to  observe the decay of the cosmic background neutrino.  
Detection of this radiative decay  means a discovery of the cosmic background neutrino 
predicted by standard cosmology.

The cosmic background neutrino has a temperature of 1.9K and a particle density $\rho$ of 
110 cm$^{-3}$ per generation.

Atmospheric neutrino data with Super-Kamiokande and neutrino beam data with K2K and MINOS 
gives the mass-square difference between $\nu_3$ and $\nu_2$,   
$\Delta m_{32}^2$ of $( 2.43 \pm 0.13 ) \times 10^{-3}eV^2$  \cite{pdg} and the mixing angle 
of sin$^{2}2\theta _{23} > $ 0.92 at 90\% confidence level \cite{pdg}.

Solar neutrino data with Super-Kamiokande, SNO and the reactor neutrino data with KamLAND 
give the mass-square difference between $\nu_2$ and $\nu_1$ of 
 $\Delta m_{12}^2 = ( 7.59 +0.19/ -0.21 ) \times 10^{-5} eV^2$ \cite{pdg} and the mixing angle 
of sin$^{2}2\theta_{12} = 0.87 \pm 0.03$ \cite{pdg}.

We search for the neutrino radiative decay under these constraints on the mass-square 
differences and the mixing angles between the neutrino generations. For this search, 
we  make a fit of the photon energy spectrum of Cosmic Infrared Background (CIB) to 
a sum of a neutrino 
decay photon energy spectrum and a continuum spectrum.

In the left-right symmetric model, we make a feasibility study of the neutrino decay search in
the CIB energy spectrum measurement.

\section{Neutrino Radiative Decay}

  In this paper  we report the analysis results on the normal hierarchy case where the $\nu_3$ 
is the heaviest
neutrino and 
decay into $\nu_2$ or $\nu_1$ as $\nu_3 \rightarrow \nu_2$ ( or $\nu_1$ ) + 
$\gamma$.

In the $\nu_{3}$ rest frame, the decay photon energy $E_{0}$ is related to the 
neutrino masses by

$E_{0} = \frac{ m_3^2 - m_2^2 }{2m_3}$.

\begin{figure}[p]
 \vspace{0.5cm}
\begin{center}
 \includegraphics[width=0.9\textwidth]{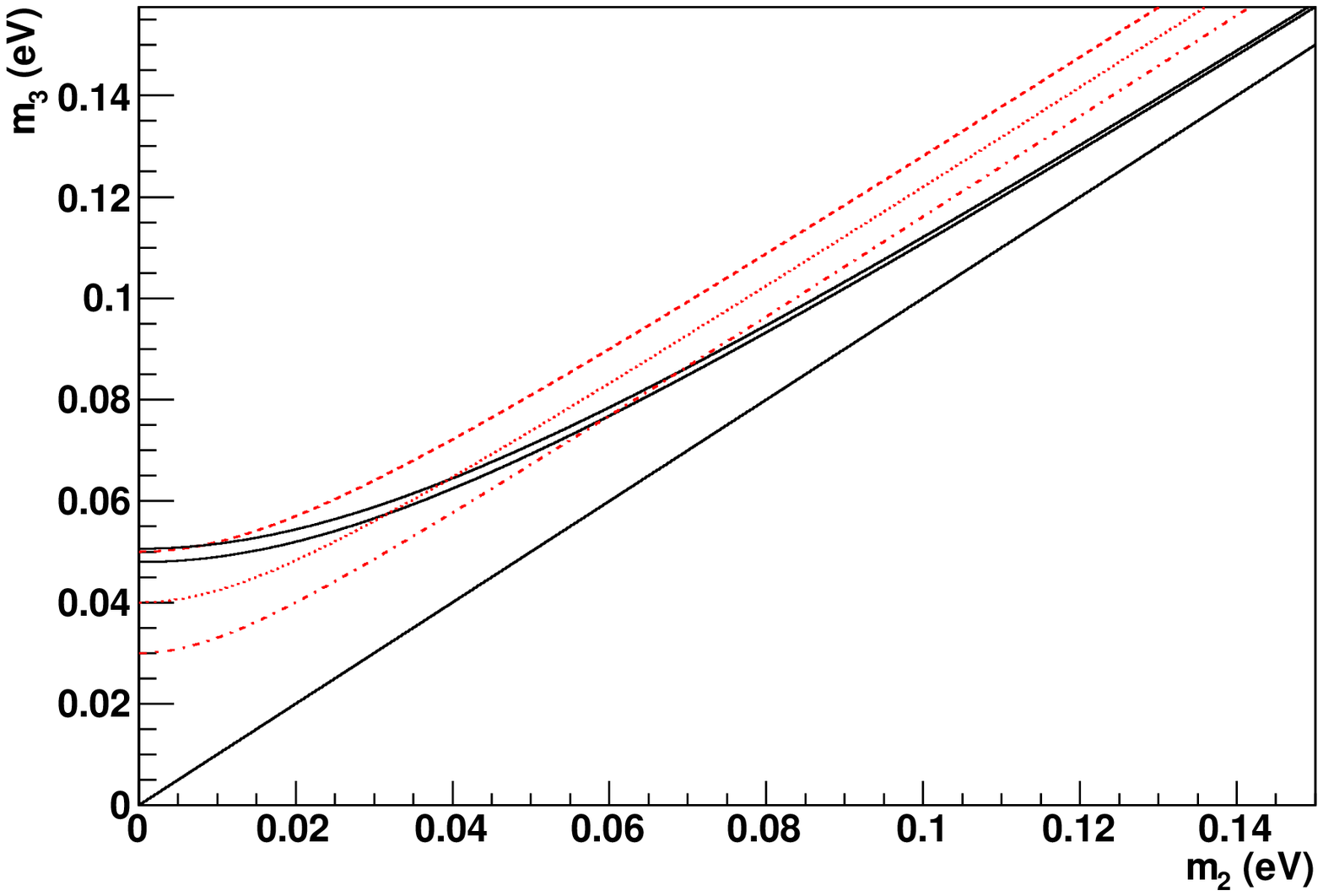}
\caption{ Relation between $m_3$ and $m_2$. The solid band shows the 1$\sigma$ constraint by 
the neutrino oscillation measurements ($\Delta m_{32}^2$ of $( 2.43 \pm 0.13 ) \times 10^{-3}eV^2$ ). 
Three curves correspond to various neutrino decay photon energies of 25meV (dashed), 20meV (dotted) 
and 15meV(dot-dashed). (Color online)}
\label{m2m3}
\end{center}
\end{figure}

\begin{figure}[p]
 \vspace{0.5cm}
\begin{center}
 \includegraphics[width=0.9\textwidth]{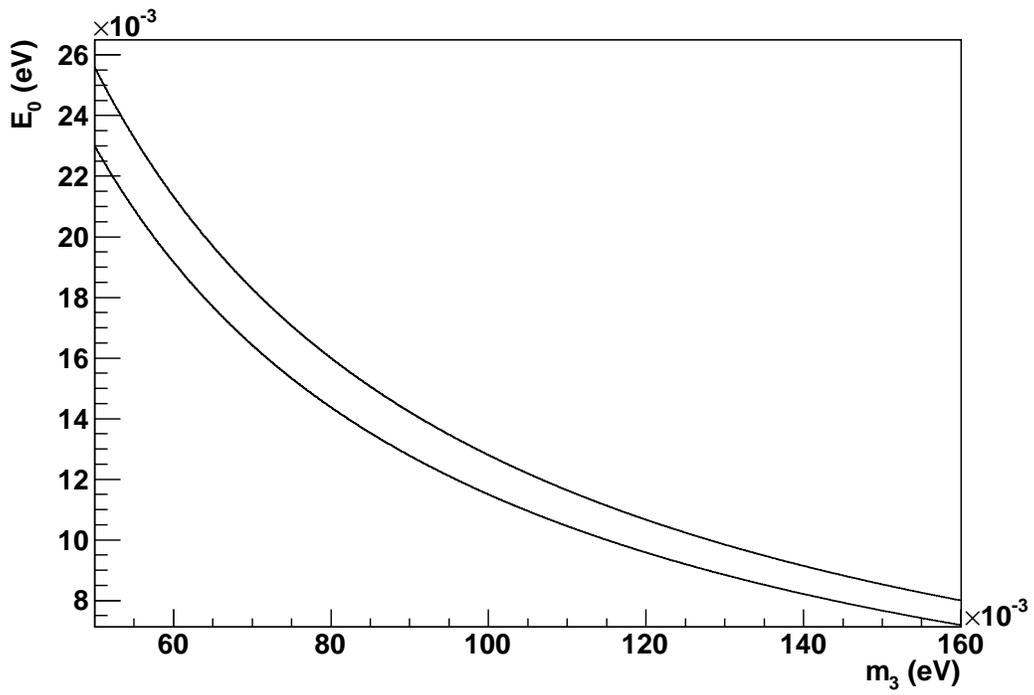}
\caption{ Relation between $E_{0}$ and $m_3$. The solid band shows the 1$\sigma$ 
constraint by the neutrino oscillation measurements ($\Delta m_{32}^2$ of $( 2.43 \pm 0.13 ) 
\times 10^{-3}eV^2$ ). }
\label{egamma-m3}
\end{center}
\end{figure}

 Since the term $m_3^2 - m_2^2$ was measured by the 
 neutrino oscillation experiments, 
we can determine $m_3$ by measuring this $E_{0}$. 
The plot of $m_3$ versus $m_2$ is 
shown in Fig.~\ref{m2m3}. The decay photon energy $E_{0}$ is given as a function 
of  $m_3$ with the neutrino oscillation measurement constraint as shown in 
Fig.~\ref{egamma-m3}.  
When we consider the $m_3$ range  between 50 \mmev\ and  150 \mmev\ ,  $E_{0}$ ranges 
from 25 meV to  8 meV in the far-infrared region.

Taking into account the redshift of the decay photon energy, the observed photon energy 
$E_{\gamma}$ is given by

$E_{\gamma} = \frac{E_{0}}{1+z}$ ,

 where z is a redshift.
As the lifetime become longer because the neutrino is moving away, the decay rate of 
neutrino R is given by
R = $\frac{1}{\tau (1+z)}$ , where $\tau$ is a neutrino lifetime at the rest frame.
Thus the decay photon flux per unit solid angle is given by

$\frac{dN_{\gamma}}{dS dt d\Omega } = \frac{ \rho }{4\pi \tau (1+z)} dr$,

where r is a distance between a neutrino decay point and a telescope 
with a detection area dS, 
and $\rho$ is a density of the cosmic background neutrino of 110 cm$^{-3}$.
Assuming a flat universe, dr is related with dz by

dr = $ \frac{ c}{H_{0}}[(1+z)^3 \Omega_{M} + \Omega_{\Lambda}]^{-0.5}dz$,

where $H_{0}$ is a Hubble constant of ( 74 $ \pm$ 4 ) km s$^{-1}$ Mpc$^{-1}$, 
and $\Omega_{M}$ and $\Omega_{\Lambda}$ are the matter density and the cosmological 
constant which were measured to be \cite{pdg}

$\Omega_{M} = 0.26 \pm 0.02$,
 
$\Omega_{\Lambda} = 0.74 \pm 0.03$,
  
$\Omega_{M} + \Omega_{\Lambda}= 1.006 \pm 0.006$.

Thus under the assumption of a flat universe with 
$\Omega_{M}$ + $\Omega_{\Lambda}$ = 1, 
we obtain the following distribution of the decay photon flux $\frac{dN_{\gamma}}{dS dt}$ 
per unit solid angle and unit energy : 

$\frac{dN_{\gamma}}{dS dt d\Omega dE_{\gamma}} = \frac{ \rho c}
{4\pi \tau H_{0}E_{\gamma}}[( \frac{E_{0}}{E_{\gamma}})^3 \Omega_{M} + \Omega_{\Lambda}]^{-0.5} $.

If we use the energy flux per unit solid angle $I$ defined by 

$ I \equiv \frac{ d( N_{\gamma} \times E_{\gamma} )}{dS dt d\Omega}$,

we obtain the energy flux per unit solid angle and unit energy given by

$ \frac{dI}{dE_{\gamma}} =  \frac{ \rho c}
{4\pi \tau H_{0}}[( \frac{E_{0}}{E_{\gamma}})^3 \Omega_{M} + \Omega_{\Lambda}]^{-0.5} $.

Using $E_{\gamma} = h\nu$ ,

$ \nu \frac{dI}{d\nu} = \frac{ \rho c}
{4\pi \tau H_{0}} h\nu [( \frac{E_{0}}{h\nu})^3 \Omega_{M} + \Omega_{\Lambda}]^{-0.5} $.

This spectrum is smeared by the neutrino motion at 1.9K, but this effect is negligibly small.  
The photon energy spreads due to this neutrino motion 
are 0.6\%  and 1.5\% in rms for the photon energies of  
25 meV and 10 meV, respectively.

\section{
 Neutrino Lifetime Limit
}

   The Cosmic Infrared Background (CIB) 
continuum in photon energy spectrum is the most serious foreground 
against this neutrino radiative decay 
signal because the sharp edge of the signal spectrum is 
located between 8 meV and 25 meV in our search region. 
There have been studies where the neutrino lifetime limit was obtained 
using the CIB data \cite{nulifetime1,nulifetime2}.
However they did not use the photon energy spectrum expected 
from the neutrino radiative decay
to estimate the neutrino lifetime limit but used only the total decay rate of neutrino. 
They did not use the AKARI CIB data which were not available 
at that time.

\begin{figure}[p]
 \vspace{0.5cm}
\begin{center}
 \includegraphics[width=0.8\textwidth]{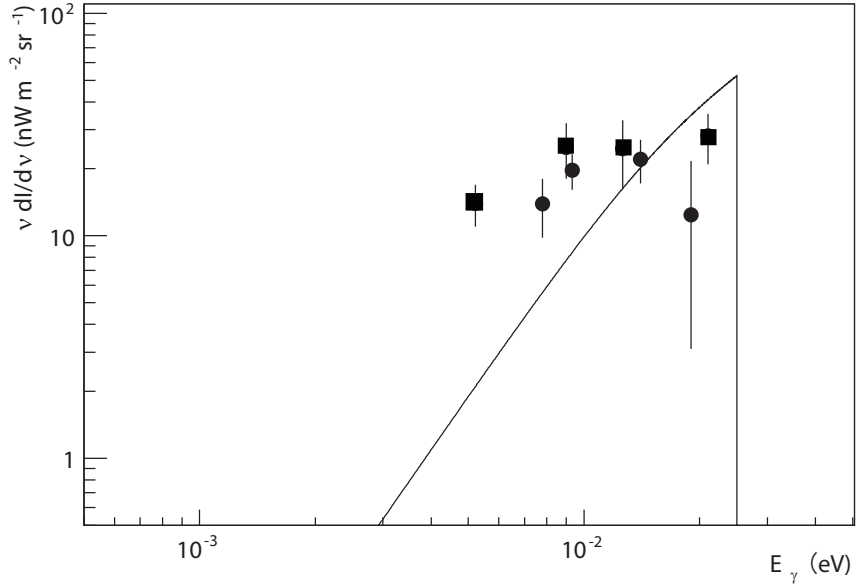}\\
 \includegraphics[width=0.8\textwidth]{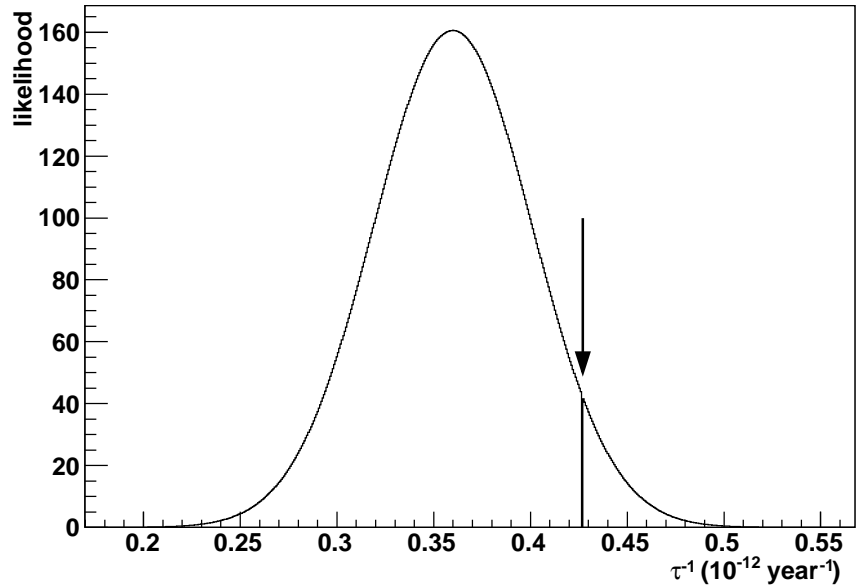}
\caption{ Top plot shows the CIB data measured by COBE (dark square) 
and AKARI (dark circle) fitted 
to the photon energy spectrum expected 
from the neutrino radiative decay for $m_3$ = 50 \mmev\ and $m_2$ = 10 \mmev\ 
with a maximum 
likelihood method. The curve shows the best fit.
Bottom plot shows the likelihood as a function of the $\nu_3$ lifetime. 
An arrow points to a  lower limit of the neutrino lifetime at 95 \% confidence level.
 }
\label{CIBfit}
\end{center}
\end{figure}

\begin{figure}[p]
 \vspace{0.5cm}
\begin{center}
 \includegraphics[width=0.9\textwidth]{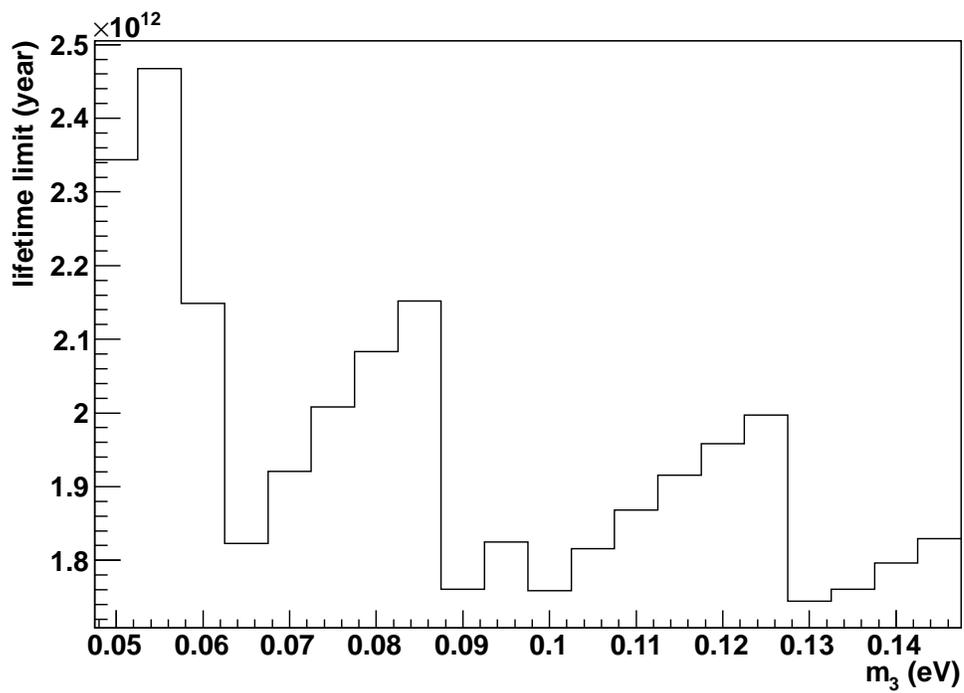}
\caption{ 
Lower limits of the neutrino lifetime  at 95 \% confidence level as a function of $m_3$ 
obtained with the CIB data measured by COBE and AKARI. }
\label{limit1}
\end{center}
\end{figure}

We perform a statistical test of the CIB data measured by 
COBE \cite{COBE1,COBE2} and AKARI \cite{AKARI} 
to estimate the neutrino lifetime limit.
 Since we do not have a sharp edge at high energy end of the CIB energy spectrum, we set 
a lower limit of the heaviest neutrino lifetime. 
The CIB data measured by COBE and AKARI were directly compared with 
the photon energy spectrum expected from the neutrino radiative decay neglecting 
the CIB continuum with a maximum likelihood method. In this
comparison, we neglected the CIB continuum in order to estimate the neutrino lifetime limit 
in the worst case which gives the lowest lifetime limit.  
The curve with a maximum likelihood  is shown together with
the likelihood as a function of 
$\nu_3$ lifetime for $m_3$ = 50 \mmev\ and $m_2$ = 10 \mmev\ in Fig~\ref{CIBfit}. 
The arrow point in this figure gives us a
lower limit of $\nu_3$ lifetime at 95\% confidence level.
We performed this procedure for various $\nu_3$ masses to obtain a  lower limit of the $\nu_3$ 
lifetime at 95\% confidence level as a function of $\nu_3$ mass as shown in 
Fig.~\ref{limit1}.  
They  range from 
$ 1.7 \times 10^{12}$  to $ 2.4 \times 10^{12}$  years 
 in the $\nu_3$ mass range between 50 \mmev\ and 150 \mmev.

 Recent deep galaxy surveys with infrared satellites,
AKARI \cite{AKARI}, Spitzer \cite{spitzer} and Hershel \cite{hershel} have revealed 
that more than a half of 
the CIB energy originates from external galaxies. 
So we performed the analysis to obtain 
stronger constraint on the neutrino lifetime taking the integrated flux of galaxies 
into account.  
We subtracted  the contribution of the distant galaxies as source points \cite{spitzer}
from the CIB measured by AKARI as shown in Table~\ref{CIBsubtraction}.
 Since there were no measured values of the
 contributions at wavelengths of 60, 90 and 140 $\mu$m, we interpolated the measured values at 
wavelengths of 24, 70 and 160 $\mu$m.
Thus we obatined the CIB after subtracting the contribution of distant galaxies.
These corrected CIB data give a 
constraint on the neutrino lifetime as the most stringent 
lower limits.

\begin{figure}[p]
 \vspace{0.5cm}
\begin{center}
 \includegraphics[width=0.8\textwidth]{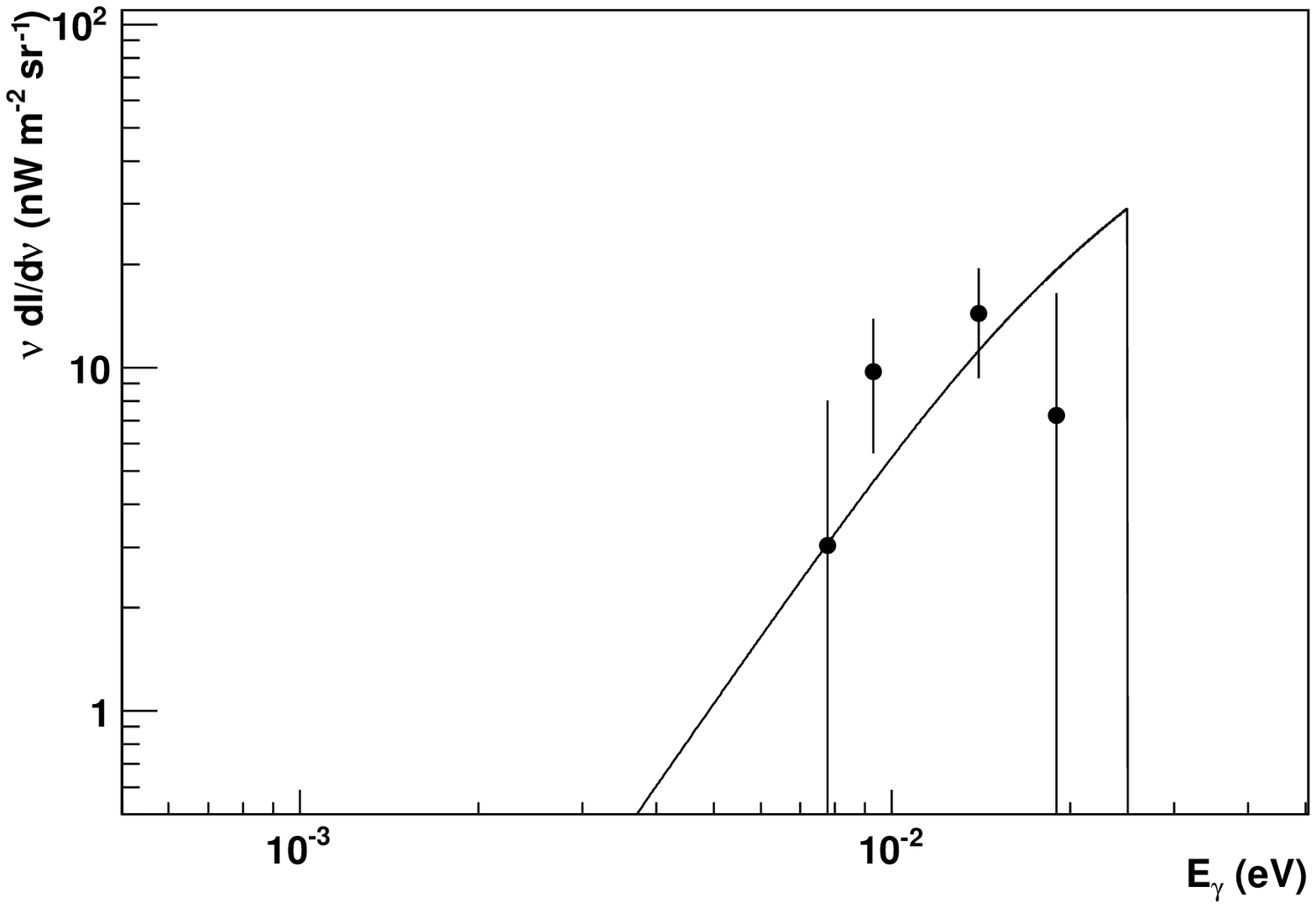}\\
 \includegraphics[width=0.8\textwidth]{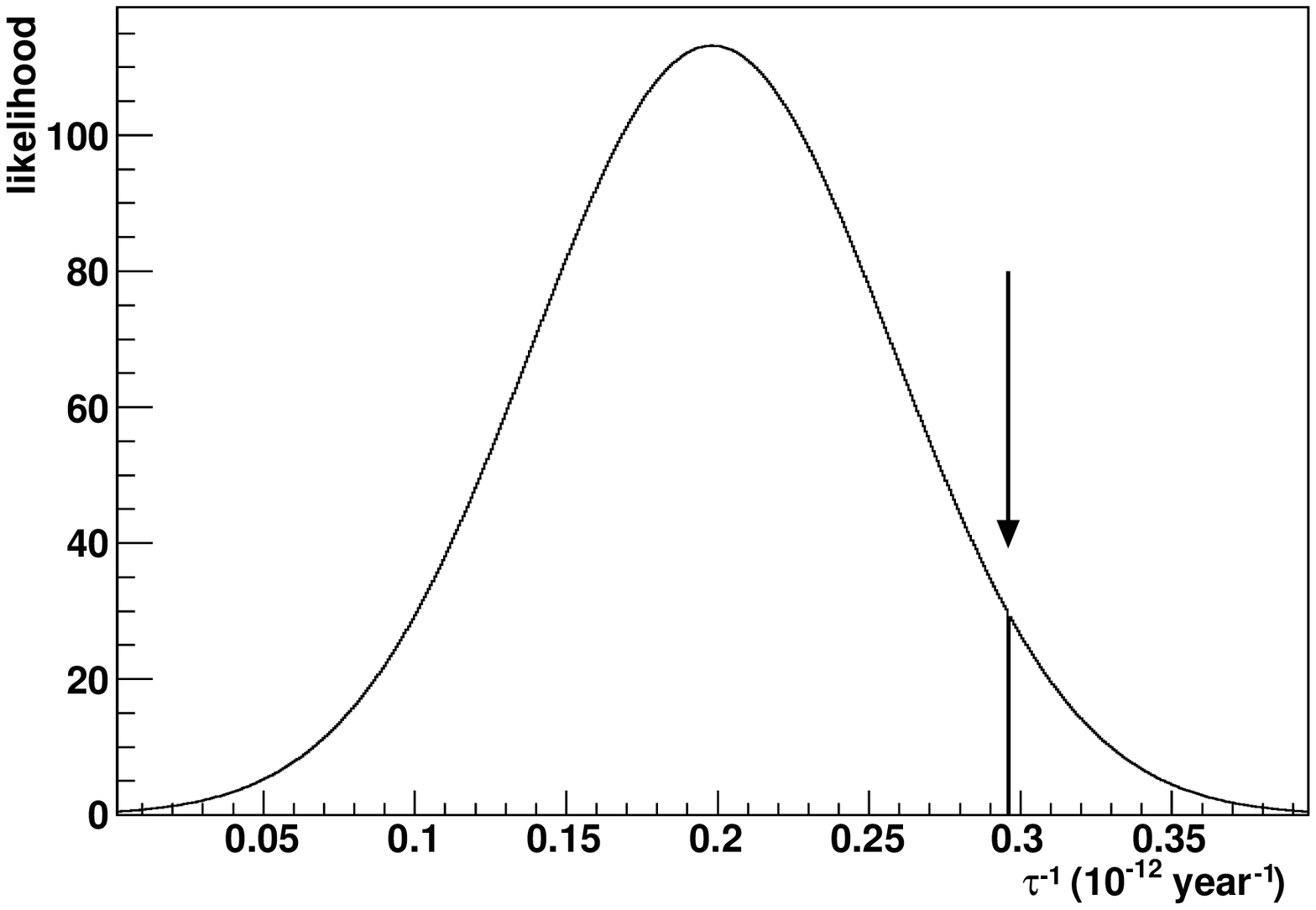} 
\caption{ Top plot 
shows the AKARI CIB data after subtracting the contribution 
of distant galaxies as point sources (dark circle) fitted 
to the photon energy spectrum expected 
from the neutrino radiative decay for $m_3$ = 50 \mmev\ and $m_2$ = 10 \mmev\ 
with a maximum 
likelihood method. The curve shows the best fit.
Bottom plot shows the likelihood as a function of $\nu_3$ lifetime. 
An arrow points to a  lower limit of the neutrino lifetime at 95 \% confidence level.
 }
\label{CIBfit-AKARI}
\end{center}
\end{figure}

\begin{figure}[p]
 \vspace{0.5cm}
\begin{center}
\includegraphics[width=0.9\textwidth]{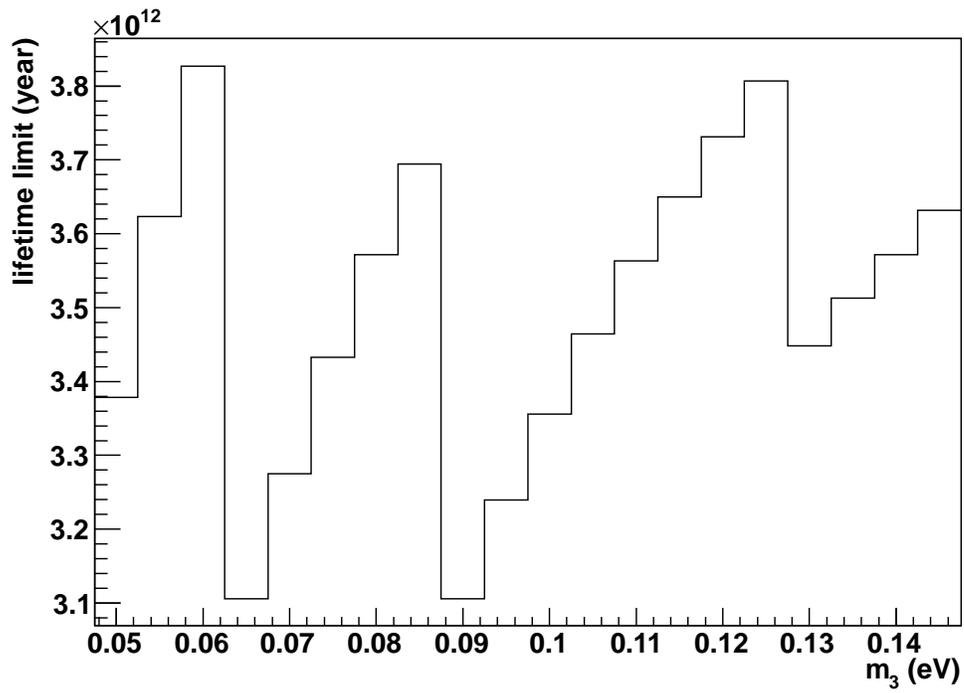}
\caption{ 
Lower limits of the neutrino lifetime  at 95 \% confidence level as a function of $m_3$ 
obtained with the CIB data measured by AKARI after subtracting the contribution of distant galaxies. }
\label{limit2}
\end{center}
\end{figure}

With the AKARI CIB data after subtracting the contributions of distant galaxies 
as point sources, we performed  a statistical test of the spectrum 
to obtain the neutrino lifetime limit, motivated by the fact that 
 such corrected AKARI CIB data has much less contributions of distant
galaxies than the COBE CIB data. 
The corrected AKARI CIB data were compared to the photon energy spectrum expected 
from the neutrino radiative decay with a maximum likelihood method. 
The curve with a maximum likelihood  is shown together with
the likelihood as a function of 
$\nu_3$ lifetime for $m_3$ = 50 \mmev\ and $m_2$ = 10 \mmev\ in Fig~\ref{CIBfit-AKARI}. 
An arrow point in this figure gives us a
lower limit of $\nu_3$ lifetime at 95\% confidence level. 
We performed this procedure for various $\nu_3$ masses to obtain a  lower limit of the $\nu_3$ 
lifetime at 95\% confidence level as a function of $\nu_3$ mass as shown in 
Fig.~\ref{limit2}.  
  They range from 3.1 $\times 10^{12}$  to $ 3.8 \times 10^{12}$  years in 
the $\nu_3$ mass range between 50 \mmev\ and 150 \mmev.

\section{Neutrino Lifetime in the Left-Right Symmetric SU(2)$_L \times$ 
SU(2)$_R \times$ U(1) Model}

   In the standard model, the heaviest neutrino lifetime is predicted to be $10^{43}$ 
year for $\nu_3$ 
with a mass of 50 \mmev \cite{SM_nudecay1,SM_nudecay2,LR_model1,LR_model2}. 
It is too long to be measured by the present method.

   In the left-right symmetric model, the lifetime is predicted to be much shorter than in 
the standard model. The left-right symmetric SU(2)$_L \times$ SU(2)$_R \times$ U(1) model 
has two 
charged weak bosons, 
or the left-handed weak boson $W_L$ and the right-handed weak boson $W_R$ which are mixed with a 
mixing angle 
$\zeta$ into two mass eigenstates $W_1$ and $W_2$ as follows \cite{LR_model1,LR_model2}:

$W_1 = W_L$cos$\zeta$ - $W_R$sin$\zeta$

$W_2 = W_L$sin$\zeta$ + $W_R$cos$\zeta$ \\ 

In this model, the lifetime of $\nu_3$ is given by

\begin{figure}[p]
 \vspace{0.5cm}
\begin{center}
 \includegraphics[width=0.9\textwidth]{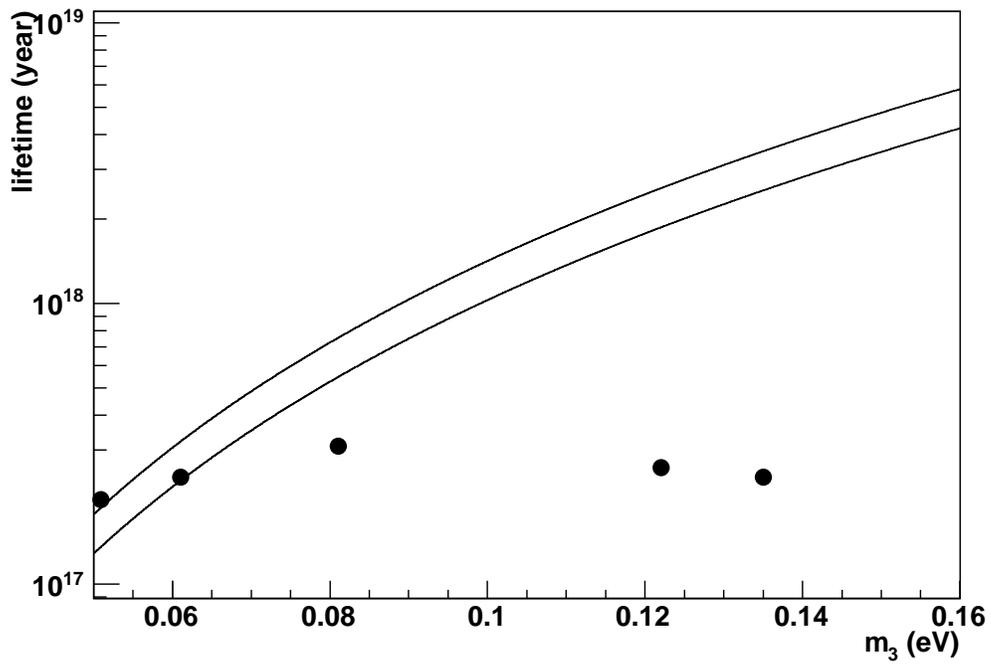}
\caption{ Expected lifetime minimum in the left-right symmetric model as a function of  $m_3$.
 The solid band shows the 1$\sigma$ constraint by the neutrino oscillation measurements 
($\Delta m_{32}^2$ of $( 2.43 \pm 0.13 ) \times 10^{-3}eV^2$ ). The sensitivity of 5 $\sigma$
observation with the proposed measurement is shown by dark circles. }
\label{lifetime-m3}
\end{center}
\end{figure}

 $\tau^{-1} = \frac{\alpha G_F^2}{128 \pi^4} (\frac{m_3^2 - m_2^2}{m_3})^3 
|U_{32}|^2 |U_{33}|^2
[ \frac{9}{64} (m_3^2 + m_2^2) \frac{m_{\tau}^4}{M_{W1}^4}(1 + \frac{M_{W1}^2}{M_{W2}^2})^2
+ 4 m_{\tau}^2 ( 1 -  \frac{M_{W1}^2}{M_{W2}^2})^2  $ sin$^2 2\zeta$ ],

where $\alpha$ is a fine structure constant, $G_F$ is a Fermi coupling constant, $m_{\tau}$,
$M_{W1}$ and  $M_{W2}$ are masses of $\tau$, $W_1$ and $W_2$, 
respectively \cite{LR_model1,LR_model2}. $U_{ij}$ is the (i, j)-th element of the Maki-Nakagawa-Sakata mixing 
matrix \cite{MNS} and we took 
$|U_{32}| = 1/\sqrt{2}$ and $|U_{33}| = 1/\sqrt{2}$.  
The present mass limit  of the right-handed weak boson $W_R$ is $M_R >$ 
0.715TeV assuming $\zeta$ = 0. The upper limit of the mixing angle between $W_L$ and $W_R$ is 
sin$\zeta <$ 0.013 \cite{pdg}.   
The lifetime is shown as a function of $m_3$ with  $M_{W_2} $ of 
0.715TeV, sin$\zeta $ of 0.013 and $\Delta m_{32}^2$ of $ (2.43 \pm 0.13 ) \times 10^{-3}eV^2$ 
in Fig.~\ref{lifetime-m3}.
  
For $m_3$ of 50 \mmev, 
the lifetime is calculated to be $1.5 \times 10^{17}$ years under the above conditions in left-right
symmetric model.  The present lifetime limit is shorter than this 
prediction by a factor of  2 $\times 10^{-5}$.

  We can improve the sensitivity of the search for neutrino radiative decay 
by measuring the continuous energy spectrum of 
the Cosmic Infrared Background (CIB). 
We fit the continuous spectrum with a sum of the CIB continuum and the neutrino decay 
photon spectrum to search for a sharp edge of the neutrino decay photon spectrum.

  To estimate the energy resolution requirement, we performed the simulation study of 
the neutrino radiative decay  assuming the following conditions:
\begin{itemize}
  \item{ The present CIB continuum is dominated by other sources than 
  the neutrino radiative decay.  This CIB spectrum is represented by a  Planck distribution
 plus a quadratic function.}
  \item{  $m_3$ = 50 \mmev\ and $m_2$ = 10 \mmev. }
  \item{  $\tau_3 = 1.5 \times 10^{17}$ years. }
\end{itemize}

\begin{figure}[p]
 \vspace{0.5cm}
\begin{center}
 \includegraphics[width=0.8\textwidth]{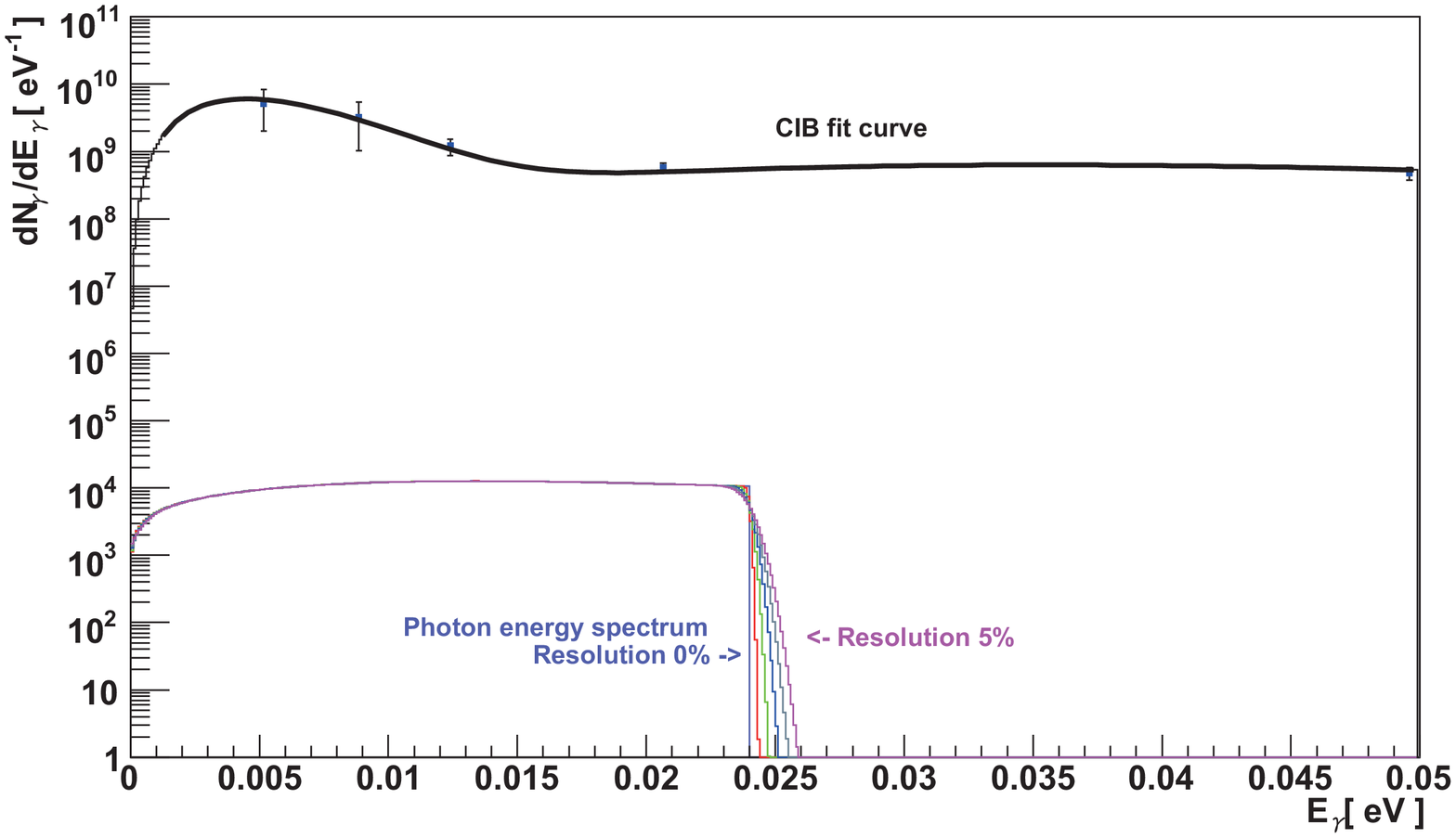}\\
 \includegraphics[width=0.8\textwidth]{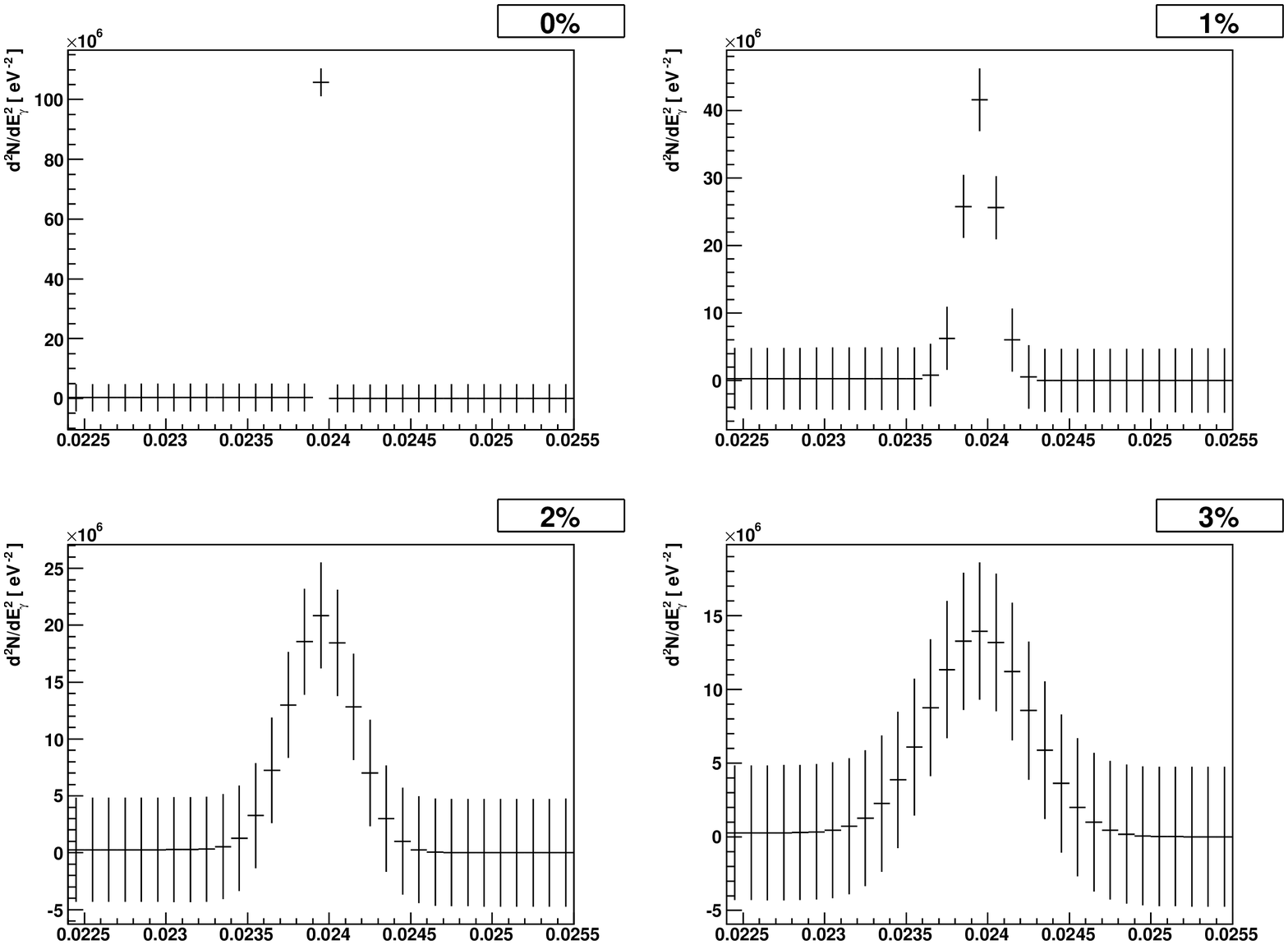}
\caption{ The top plot shows the photon energy spectra of the CIB and the neutrino 
radiative decay with
various energy  resolutions from 0 \% to 5 \% by 1 \% step. 
The CIB continuum is fitted to a sum of a function of the Planck 
distribution and a quadratic function. 
The bottom plot is the negative energy derivative of the photon energy 
spectrum for a sum of the CIB continuum and the neutrino radiative decay
for energy resolutions of 0, 1, 2 and 3 \%. (Color online)}
\label{spectrum-sim}
\end{center}
\end{figure}

 We simulated the CIB detection experiments with a 20cm-diameter telescope with a viewing angle 
of 0.1 degrees and energy resolutions of 
0 to 5 \% by 1 \% step.
 With 100\% detection efficiency and 
10-hour data taking, we have the photon energy spectrum for the CIB and the neutrino radiative 
decay
as shown in Fig.~\ref{spectrum-sim}. The CIB continuum was fitted to a function of the Planck 
distribution plus a quadratic function. Then we calculated the energy derivative of the photon 
energy spectrum for a sum of the CIB  continuum 
and the neutrino radiative decay as shown in  Fig.~\ref{spectrum-sim}.
In this negative energy derivative plot, the sharp edge of the energy spectrum is identified as 
a clean peak.
The excess is 6.7$\sigma$ and we can see the clear signal peak in the distribution 
if the energy resolution is less than 2\%. 

 We search for the neutrino radiative decay in the $m_3$ range between 50 \mmev\ 
and  150 \mmev\ which 
corresponds to the photon energy  range between 25 meV and  8 meV with 
this experiment. Thus the photon energy range to be measured is between 5 meV  
( $\lambda$ = 250 $\mu$m ) 
and 35 meV ( $\lambda$ = 35 $\mu$m). In this photon energy range, the frequency of 
photon coming in 
this telescope is expected to be around 5 MHz. As we will use 400 pixels as a photon detector, 
the photon
rate per pixel is around 12 kHz/pixel.

By this simulation study, we found that the required energy resolution is less than 2\% at 
25 meV.  With this resolution, we estimated 5$\sigma$ observation lifetime with  10-hour running 
of this telescope from the differential CIB energy spectrum including the radiative decay 
of the cosmic background neutrino for 
the $m_3$ range between 50 \mmev\ and  150 \mmev\
 with $\Delta m_{32}^2$ of $  2.43 \times 10^{-3}eV^2$. 
 The 5$\sigma$ observation lifetime ranges from $2.0 \times 
10^{17}$ to $3.0 \times 10^{17}$ 
years as shown in Fig.~\ref{lifetime-m3}.
Spectral emission features of the dusts in galaxies at redshifts of 1, when a large fraction 
of the CIB energy was generated \cite{spitzer}, may produce a structure in the spectral energy
distribution (SED) of CIB, similar to the neutrino decay photon SED.
However, SEDs of such distant galaxies in the energy range where the neutrino decay photon is 
expected to be observed, have not been well explored. Further study of the SED of 
distant galaxies
with future, large-aperture, infrared telescope
such as JAXA's SPICA (Space Infrared Telescope for Cosmology and Astrophysics) 
mission \cite{spica} is important to mitigate the spectral contamination
of the neutrino decay photon.

The present measured CIB spectrum includes the point-sources of distant 
galaxies and the zodiacal light 
foreground ambiguity. By using small pixels with small viewing angle such as 0.005 degrees, 
we will be able to 
distinguish the distant galaxy point sources from the CIB spectrum. In the future project 
EXZIT ( Exo-Zodiacal Infrared Telescope ) \cite{exzit}
which will observe the CIB outside of Jupiter orbit, we will decrease the effect of the 
zodiacal emission
significantly, and will have much less ambiguity of the zodiacal light foreground.

\section{Conclusion}

        We propose to search for the neutrino radiative decay by fitting 
a photon energy spectrum of the cosmic infrared background 
to a sum of the photon energy spectrum from the neutrino radiative decay and a continuum. 
By comparing the present cosmic infrared background energy spectrum observed by AKARI 
and Spitzer to the photon energy spectrum expected from neutrino radiative decay 
with a maximum likelihood method, we obatined a lower limit of netrino lifetime  
 of 3.1 $\times 10^{12}$  to 3.8  $\times 10^{12}$ years 
at 95 \% confidence level for $\nu_3$  in the mass range
between 50 \mmev\ and 150 \mmev\ under the present constraints by neutrino oscillation 
measurements.

   In the left-right symmetric model, the lifetime is predicted to be much shorter than in 
the standard model. The present mass limit  of the right-handed weak boson $W_R$ is $m_R >$ 
0.715 TeV and the mixing angle between $W_L$ and $W_R$ is sin$\zeta <$ 0.013 .  In this model, 
the minimum lifetime of $\nu_3$ is predicted to be
 $1.5 \times 10^{17}$ years  for $m_3$ of 50 \mmev. 

The present lifetime limit is shorter than this 
prediction by  a factor of 2 $\times$ 
$10^{-5}$. By measuring the continuous energy spectrum of the cosmic infrared
 background  in the photon energy region between 5 meV ($\lambda$ = 250 $\mu$m) 
and 35 meV ($\lambda$ = 35 $\mu$m), we  can expect  5$\sigma$ 
observation of the radiative decay 
of neutrino with a lifetime of $1.5 \times 10^{17}$ years.\\

\begin{Large}
\bf{Acknowlegements}\\
\end{Large}

        We thank M. Yamauchi and Y. Okada at KEK for discussions regarding the neutrino decay and 
its lifetime calculation.
This work was supported by the Ministry of Education, Science, Sports and Culture of Japan.
Part of this work was supported by KAKENHI(19540250 and 21111004).
 

{
}

\begin{table}[hp]
\begin{tabular}{c|c|c|c|c}
\hline
Wavelength & CIB$^a$ & contribution of  & 
CIB  &  CIB  \\
& & distant galaxies$^b$ & after subtarction$^c$ & after subtraction$^d$ \\
($\mu$m) & (MJy sr$^{-1}$) & (MJy sr$^{-1}$) & (MJy sr$^{-1}$) & (nW m$^{-2}$ sr$^{-1}$)\\
\hline
24 &                  & 0.017 $\pm$ 0.003     &                   & \\
60 & 0.27 $\pm$ 0.20  & ( 0.112 $\pm$ 0.018 ) & 0.16 $\pm$ 0.20   & 7.3 $\pm$ 9.2 \\
70 &                  & 0.138 $\pm$ 0.023     &                   & \\
90 & 0.67 $\pm$ 0.15  & ( 0.234 $\pm$ 0.042 ) & 0.44 $\pm$ 0.16   & 14.4 $\pm$ 5.1 \\
140 & 0.94 $\pm$ 0.17 & ( 0.475 $\pm$ 0.095 ) & 0.47 $\pm$ 0.20   & 9.8 $\pm$ 4.1  \\
160 & 0.73 $\pm$ 0.21 & 0.571 $\pm$ 0.120     & 0.16 $\pm$ 0.24   & 3.0 $\pm$ 5.0 \\
\hline
\end{tabular}
\caption
{ The AKARI CIB data before and after subtracting the contribution of distant galaxies.\newline
$^a$ The CIB measured by AKARI in unit of MJy sr$^{-1}$.\newline
$^b$ The contribution of distant galaxies  to the CIB measured by Spitzer 
in unit of MJy sr$^{-1}$. The numbers in parentheses were obtained by a linear interpolation.\newline
$^c$ The CIB measured by AKARI after subtracting the contribution of distant galaxies 
in unit of MJy sr$^{-1}$.\newline
$^d$ The CIB measured by AKARI after subtracting the contribution of distant galaxies
 in unit of nW m$^{-2}$ sr$^{-1}$.}
\label
{CIBsubtraction}
\end{table}


\normalsize
\begin{thebibliography}{9}

\bibitem{SK1}
Y. Fukuda, T. Hayakawa, E. Ichihara, K. Inoue, K. Ishihara, H. Ishino, Y. Itow, T. Kajita, 
J. Kameda, S. Kasuga, K. Kobayashi, Y. Kobayashi, Y. Koshio, M. Miura, M. Nakahata, 
S. Nakayama, A. Okada, K. Okumura, N. Sakurai, M. Shiozawa, Y. Suzuki, Y. Takeuchi, Y. Totsuka, 
S. Yamada, M. Earl, A. Habig, E. Kearns, M. D. Messier, K. Scholberg, J. L. Stone, 
L. R. Sulak, C. W. Walter, M. Goldhaber, T. Barszczxak, D. Casper, W. Gajewski, 
P. G. Halverson, J. Hsu, W. R. Kropp, L. R. Price, F. Reines, M. Smy, H. W. Sobel, 
M. R. Vagins, K. S. Ganezer, W. E. Keig, R. W. Ellsworth, S. Tasaka, J. W. Flanagan, 
A. Kibayashi, J. G. Learned, S. Matsuno, V. J. Stenger, D. Takemori, T. Ishii, J. Kanzaki, 
T. Kobayashi, S. Mine, K. Nakamura, K. Nishikawa, Y. Oyama, A. Sakai, M. Sakuda, O. Sasaki, 
S. Echigo, M. Kohama, A. T. Suzuki, T. J. Haines, E. Blaufuss, B. K. Kim, R. Sanford, 
R. Svoboda, M. L. Chen, Z. Conner, J. A. Goodman, G. W. Sullivan, J. Hill, C. K. Jung, 
K. Martens, C. Mauger, C. McGrew, E. Sharkey, B. Viren, C. Yanagisawa, W. Doki, K. Miyano, 
H. Okazawa, C. Saji, M. Takahata, Y. Nagashima, M. Takita, T. Yamaguchi, M. Yoshida, 
S. B. Kim, M. Etoh, K. Fujita, A. Hasegawa, T. Hasegawa, S. Hatakeyama, T. Iwamoto, M. Koga, 
T. Maruyama, H. Ogawa, J. Shirai, A. Suzuki, F. Tsushima, M. Koshiba, M. Nemoto, K. Nishijima, 
T. Futagami, Y. Hayato, Y. Kanaya, K. Kaneyuki, Y. Watanabe, D. Kielczewska, R. A. Doyle, 
J. S. George, A. L. Stachyra, L. L. Wai, R. J. Wilkes, and K. K. Young
 (Super Kamiokande Collaboration): Phys. Rev. Lett. 81 (1998) 1562. 
\bibitem{SK2}
Y. Ashie, J. Hosaka, K. Ishihara, Y. Itow, J. Kameda, Y. Koshio, A. Minamino, C. Mitsuda, 
M. Miura, S. Moriyama, M. Nakahata, T. Namba, R. Nambu, Y. Obayashi, M. Shiozawa, Y. Suzuki,
 Y. Takeuchi, K. Taki, S. Yamada, M. Ishitsuka, T. Kajita, K. Kaneyuki, S. Nakayama,
 A. Okada, K. Okumura, T. Ooyabu, C. Saji, Y. Takenaga, S. Desai, E. Kearns, S. Likhoded,
 J. L. Stone, L. R. Sulak, C. W. Walter, W. Wang, M. Goldhaber, D. Casper, J. P. Cravens, 
W. Gajewski, W. R. Kropp, D. W. Liu, S. Mine, M. B. Smy, H. W. Sobel, C. W. Sterner, 
M. R. Vagins, K. S. Ganezer, J. Hill, W. E. Keig, J. S. Jang, J. Y. Kim, I. T. Lim, 
R. W. Ellsworth, S. Tasaka, G. Guillian, A. Kibayashi, J. G. Learned, S. Matsuno,
 D. Takemori, M. D. Messier, Y. Hayato, A. K. Ichikawa, T. Ishida, T. Ishii, T. Iwashita, 
T. Kobayashi, T. Maruyama, K. Nakamura, K. Nitta, Y. Oyama, M. Sakuda, Y. Totsuka, 
A. T. Suzuki, M. Hasegawa, K. Hayashi, T. Inagaki, I. Kato, H. Maesaka, T. Morita, 
T. Nakaya, K. Nishikawa, T. Sasaki, S. Ueda, S. Yamamoto, T. J. Haines, S. Dazeley,
 S. Hatakeyama, R. Svoboda, E. Blaufuss, J. A. Goodman, G. W. Sullivan, D. Turcan,
 K. Scholberg, A. Habig, Y. Fukuda, C. K. Jung, T. Kato, K. Kobayashi, M. Malek, 
C. Mauger, C. McGrew, A. Sarrat, E. Sharkey, C. Yanagisawa, T. Toshito, K. Miyano,
 N. Tamura, J. Ishii, Y. Kuno, Y. Nagashima, M. Takita, M. Yoshida, S. B. Kim, J. Yoo,
 H. Okazawa, T. Ishizuka, Y. Choi, H. K. Seo, Y. Gando, T. Hasegawa, K. Inoue, J. Shirai,
 A. Suzuki, M. Koshiba, Y. Nakajima, K. Nishijima, T. Harada, H. Ishino, R. Nishimura,
 Y. Watanabe, D. Kielczewska, J. Zalipska, H. G. Berns, R. Gran, K. K. Shiraishi, 
 A. Stachyra, K. Washburn, and R. J. Wilkes (Super Kamiokande Collaboration): 
 Phys. Rev. Lett. 93 (2004) 101801. 
\bibitem{K2K}
M. H. Ahn, E. Aliu, S. Andringa, S. Aoki, Y. Aoyama, J. Argyriades, K. Asakura, R. Ashie,
 F. Berghaus, H. G. Berns, H. Bhang, A. Blondel, S. Borghi, J. Bouchez, S. C. Boyd,
 J. Burguet-Castell, D. Casper, J. Catala, C. Cavata, A. Cervera, S. M. Chen, K. O. Cho,
 J. H. Choi, U. Dore, S. Echigo, X. Espinal, M. Fechner, E. Fernandez, K. Fujii, Y. Fujii,
 S. Fukuda, Y. Fukuda, J. Gomez-Cadenas, R. Gran, T. Hara, M. Hasegawa, T. Hasegawa,
 K. Hayashi, Y. Hayato, R. L. Helmer, I. Higuchi, J. Hill, K. Hiraide, E. Hirose, J. Hosaka,
 A. K. Ichikawa, M. Ieiri, M. Iinuma, A. Ikeda, T. Inagaki, T. Ishida, K. Ishihara, H. Ishii,
 T. Ishii, H. Ishino, M. Ishitsuka, Y. Itow, T. Iwashita, H. I. Jang, J. S. Jang, E. J. Jeon,
 I. S. Jeong, K. K. Joo, G. Jover, C. K. Jung, T. Kajita, J. Kameda, K. Kaneyuki, B. H. Kang,
 I. Kato, Y. Kato, E. Kearns, D. Kerr, C. O. Kim, M. Khabibullin, A. Khotjantsev,
 D. Kielczewska, B. J. Kim, H. I. Kim, J. H. Kim, J. Y. Kim, S. B. Kim, M. Kitamura,
 P. Kitching, K. Kobayashi, T. Kobayashi, M. Kohama, A. Konaka, Y. Koshio, W. Kropp,
 J. Kubota, Yu. Kudenko, G. Kume, Y. Kuno, Y. Kurimoto, T. Kutter, J. Learned,
 S. Likhoded, I. T. Lim, S. H. Lim, P. F. Loverre, L. Ludovici, H. Maesaka, J. Mallet,
 C. Mariani, K. Martens, T. Maruyama, S. Matsuno, V. Matveev, C. Mauger,
 K. B. McConnel Mahn, C. McGrew, S. Mikheyev, M. Minakawa, A. Minamino, S. Mine,
 O. Mineev, C. Mitsuda, G. Mitsuka, M. Miura, Y. Moriguchi, T. Morita, S. Moriyama,
 T. Nakadaira, M. Nakahata, K. Nakamura, I. Nakano, F. Nakata, T. Nakaya, S. Nakayama,
 T. Namba, R. Nambu, S. Nawang, K. Nishikawa, H. Nishino, S. Nishiyama, K. Nitta, S. Noda,
 H. Noumi, F. Nova, P. Novella, Y. Obayashi, A. Okada, K. Okumura, M. Okumura, M. Onchi,
 T. Ooyabu, S. M. Oser, T. Otaki, Y. Oyama, M. Y. Pac, H. Park, F. Pierre, A. Rodriguez,
 C. Saji, A. Sakai, M. Sakuda, N. Sakurai, F. Sanchez, A. Sarrat, T. Sasaki, H. Sato,
 K. Sato, K. Scholberg, R. Schroeter, M. Sekiguchi, E. Seo, E. Sharkey, A. Shima,
 M. Shiozawa, K. Shiraishi, G. Sitjes, M. Smy, H. So, H. Sobel, M. Sorel, J. Stone,
 L. Sulak, Y. Suga, A. Suzuki, Y. Suzuki, Y. Suzuki, M. Tada, T. Takahashi, M. Takasaki,
 M. Takatsuki, Y. Takenaga, K. Takenaka, H. Takeuchi, Y. Takeuchi, K. Taki, Y. Takubo,
 N. Tamura, H. Tanaka, K. Tanaka, M. Tanaka, Y. Tanaka, K. Tashiro, R. Terri, S. T. Jampens,
 A. Tornero-Lopez, T. Toshito, Y. Totsuka, S. Ueda, M. Vagins, L. Whitehead, C. W. Walter,
 W. Wang, R. J. Wilkes, S. Yamada, Y. Yamada, S. Yamamoto, Y. Yamanoi, C. Yanagisawa,
 N. Yershov, H. Yokoyama, M. Yokoyama, J. Yoo, M. Yoshida,
 and J. Zalipska  (K2K Collaboration): Phys. Rev. D74 (2006) 072003. 

\bibitem{MINOS1} 
D. G. Michael, P. Adamson, T. Alexopoulos, W. W. M. Allison, G. J. Alner, K. Anderson,
 C. Andreopoulos, M. Andrews, R. Andrews, K. E. Arms, R. Armstrong, C. Arroyo, 
D. J. Auty, S. Avvakumov, D. S. Ayres, B. Baller, B. Barish, M. A. Barker,
 P. D. Barnes, Jr., G. Barr, W. L. Barrett, E. Beall, B. R. Becker, A. Belias,
 T. Bergfeld, R. H. Bernstein, D. Bhattacharya, M. Bishai, A. Blake, V. Bocean,
 B. Bock, G. J. Bock, J. Boehm, D. J. Boehnlein, D. Bogert, P. M. Border, C. Bower,
 S. Boyd, E. Buckley-Geer, C. Bungau, A. Byon-Wagner, A. Cabrera, J. D. Chapman,
 T. R. Chase, D. Cherdack, S. K. Chernichenko, S. Childress, B. C. Choudhary,
 J. H. Cobb, J. D. Cossairt, H. Courant, D. A. Crane, A. J. Culling, J. W. Dawson,
 J. K. de Jong, D. M. DeMuth, A. De Santo, M. Dierckxsens, M. V. Diwan, M. Dorman,
 G. Drake, D. Drakoulakos, R. Ducar, T. Durkin, A. R. Erwin, C. O. Escobar,
 J. J. Evans, O. D. Fackler, E. Falk Harris, G. J. Feldman, N. Felt, T. H. Fields,
 R. Ford, M. V. Frohne, H. R. Gallagher, M. Gebhard, G. A. Giurgiu, A. Godley,
 J. Gogos, M. C. Goodman, Yu. Gornushkin, P. Gouffon, R. Gran, E. Grashorn, N. Grossman,
 J. J. Grudzinski, K. Grzelak, V. Guarino, A. Habig, R. Halsall, J. Hanson, D. Harris,
 P. G. Harris, J. Hartnell, E. P. Hartouni, R. Hatcher, K. Heller, N. Hill, Y. Ho,
 A. Holin, C. Howcroft, J. Hylen, M. Ignatenko, D. Indurthy, G. M. Irwin,
 M. Ishitsuka, D. E. Jaffe, C. James, L. Jenner, D. Jensen, T. Joffe-Minor, T. Kafka,
 H. J. Kang, S. M. S. Kasahara, J. Kilmer, H. Kim, M. S. Kim, G. Koizumi, S. Kopp,
 M. Kordosky, D. J. Koskinen, M. Kostin, S. K. Kotelnikov, D. A. Krakauer,
 A. Kreymer, S. Kumaratunga, A. S. Ladran, K. Lang, C. Laughton, A. Lebedev, R. Lee,
 W. Y. Lee, M. A. Libkind, J. Ling, J. Liu, P. J. Litchfield, R. P. Litchfield,
 N. P. Longley, P. Lucas, W. Luebke, S. Madani, E. Maher, V. Makeev, W. A. Mann,
 A. Marchionni, A. D. Marino, M. L. Marshak, J. S. Marshall, N. Mayer, J. McDonald,
 A. M. McGowan, J. R. Meier, G. I. Merzon, M. D. Messier, R. H. Milburn, J. L. Miller,
 W. H. Miller, S. R. Mishra, A. Mislivec, P. S. Miyagawa, C. D. Moore, J. Morfi'n,
 R. Morse, L. Mualem, S. Mufson, S. Murgia, M. J. Murtagh, J. Musser, D. Naples,
 C. Nelson, J. K. Nelson, H. B. Newman, F. Nezrick, R. J. Nichol, T. C. Nicholls,
 J. P. Ochoa-Ricoux, J. Oliver, W. P. Oliver, V. A. Onuchin, T. Osiecki, R. Ospanov,
 J. Paley, V. Paolone, A. Para, T. Patzak, Z. Pavlovic, G. F. Pearce, N. Pearson,
 C. W. Peck, C. Perry, E. A. Peterson, D. A. Petyt, H. Ping, R. Piteira, R. Pittam,
 A. Pla-Dalmau, R. K. Plunkett, L. E. Price, M. Proga, D. R. Pushka, D. Rahman,
 R. A. Rameika, T. M. Raufer, A. L. Read, B. Rebel, J. Reichenbacher, D. E. Reyna,
 C. Rosenfeld, H. A. Rubin, K. Ruddick, V. A. Ryabov, R. Saakyan, M. C. Sanchez,
 N. Saoulidou, J. Schneps, P. V. Schoessow, P. Schreiner, R. Schwienhorst,
 V. K. Semenov, S.-M. Seun, P. Shanahan, P. D. Shield, W. Smart, V. Smirnitsky, C. Smith,
 P. N. Smith, A. Sousa, B. Speakman, P. Stamoulis, A. Stefanik, P. Sullivan, J. M. Swan,
 P. A. Symes, N. Tagg, R. L. Talaga, A. Terekhov, E. Tetteh-Lartey, J. Thomas,
 J. Thompson, M. A. Thomson, J. L. Thron, G. Tinti, R. Trendler, J. Trevor, I. Trostin,
 V. A. Tsarev, G. Tzanakos, J. Urheim, P. Vahle, M. Vakili, K. Vaziri, C. Velissaris,
 V. Verebryusov, B. Viren, L. Wai, C. P. Ward, D. R. Ward, M. Watabe, A. Weber,
 R. C. Webb, A. Wehmann, N. West, C. White, R. F. White, S. G. Wojcicki, D. M. Wright,
 Q. K. Wu, W. G. Yan, T. Yang, F. X. Yumiceva, J. C. Yun, H. Zheng, M. Zois, 
 and R. Zwaska1 (MINOS Collaboration): Phys. Rev. Lett. 97 (2006) 191801.

\bibitem{MINOS2} 
P. Adamson, C. Andreopoulos, K. E. Arms, R. Armstrong, D. J. Auty, D. S. Ayres, B. Baller,
 P. D. Barnes, Jr., G. Barr, W. L. Barrett, B. R. Becker, A. Belias, R. H. Bernstein,
 D. Bhattacharya, M. Bishai, A. Blake, G. J. Bock, J. Boehm, D. J. Boehnlein, D. Bogert,
 C. Bower, E. Buckley-Geer, S. Cavanaugh, J. D. Chapman, D. Cherdack, S. Childress,
 B. C. Choudhary, J. H. Cobb, S. J. Coleman, A. J. Culling, J. K. de Jong,
 M. Dierckxsens, M. V. Diwan, M. Dorman, S. A. Dytman, C. O. Escobar, J. J. Evans,
 E. Falk Harris, G. J. Feldman, M. V. Frohne, H. R. Gallagher, A. Godley, M. C. Goodman,
 P. Gouffon, R. Gran, E. W. Grashorn, N. Grossman, K. Grzelak, A. Habig, D. Harris,
 P. G. Harris, J. Hartnell, R. Hatcher, K. Heller, A. Himmel, A. Holin, J. Hylen,
 G. M. Irwin, M. Ishitsuka, D. E. Jaffe, C. James, D. Jensen, T. Kafka, S. M. S. Kasahara,
 J. J. Kim, M. S. Kim, G. Koizumi, S. Kopp, M. Kordosky, D. J. Koskinen, S. K. Kotelnikov,
 A. Kreymer, S. Kumaratunga, K. Lang, J. Ling, P. J. Litchfield, R. P. Litchfield,
 L. Loiacono, P. Lucas, J. Ma, W. A. Mann, A. Marchionni, M. L. Marshak, J. S. Marshall,
 N. Mayer, A. M. McGowan, J. R. Meier, G. I. Merzon, M. D. Messier, C. J. Metelko,
 D. G. Michael, J. L. Miller, W. H. Miller, S. R. Mishra, C. D. Moore, J. Morfi'n,
 L. Mualem, S. Mufson, S. Murgia, J. Musser, D. Naples, J. K. Nelson, H. B. Newman,
 R. J. Nichol, T. C. Nicholls, J. P. Ochoa-Ricoux, W. P. Oliver, R. Ospanov,
 J. Paley, V. Paolone, A. Para, T. Patzak, Z. Pavlovic', G. Pawloski, G. F. Pearce,
 C. W. Peck, E. A. Peterson, D. A. Petyt, R. Pittam, R. K. Plunkett, A. Rahaman,
 R. A. Rameika, T. M. Raufer, B. Rebel, J. Reichenbacher, P. A. Rodrigues, C. Rosenfeld,
 H. A. Rubin, K. Ruddick, V. A. Ryabov, M. C. Sanchez, N. Saoulidou, J. Schneps,
 P. Schreiner, S.-M. Seun, P. Shanahan, W. Smart, C. Smith, A. Sousa, B. Speakman,
 P. Stamoulis, M. Strait, P. Symes, N. Tagg, R. L. Talaga, M. A. Tavera, J. Thomas,
 J. Thompson, M. A. Thomson, J. L. Thron, G. Tinti, I. Trostin, V. A. Tsarev,
 G. Tzanakos, J. Urheim, P. Vahle, B. Viren, C. P. Ward, D. R. Ward, M. Watabe,
 A. Weber, R. C. Webb, A. Wehmann, N. West, C. White, S. G. Wojcicki, D. M. Wright,
 T. Yang, M. Zois, K. Zhang, and R. Zwaska  (MINOS Collaboration): 
 Phys. Rev. Lett. 101 (2008) 131802. 
\bibitem{solar-SK} 
 S. Fukuda, Y. Fukuda, M. Ishitsuka, Y. Itow, T. Kajita, J. Kameda, K. Kaneyuki, K. Kobayashi,
 Y. Koshio, M. Miura, S. Moriyama, M. Nakahata, S. Nakayama, T. Namba, A. Okada, N. Sakurai,
 M. Shiozawa, Y. Suzuki, H. Takeuchi, Y. Takeuchi, Y. Totsuka, S. Yamada, S. Desai, M. Earl,
 E. Kearns, M.D. Messier, 1, J.L. Stone, L.R. Sulak, C.W. Walter, M. Goldhaber, T. Barszczak,
 D. Casper, W. Gajewski, W.R. Kropp, S. Mine, D.W. Liu, M.B. Smy, H.W. Sobel, M.R. Vagins,
 A. Gago, K.S. Ganezer, W.E. Keig, R.W. Ellsworth, S. Tasaka, A. Kibayashi, J.G. Learned,
 S. Matsuno, D. Takemori, Y. Hayato, T. Ishii, T. Kobayashi, T. Maruyama,  K. Nakamura,
 Y. Obayashi, Y. Oyama, M. Sakuda, M. Yoshida, M. Kohama, T. Iwashita, A.T. Suzuki,
 A. Ichikawa, T. Inagaki, I. Kato, T. Nakaya, K. Nishikawa, T.J. Haines, d, S. Dazeley,
 S. Hatakeyama, R. Svoboda, E. Blaufuss, M.L. Chen, J.A. Goodman, G. Guillian, G.W. Sullivan,
 D. Turc, K. Scholberg, A. Habig, M. Ackermann, J. Hill, C.K. Jung, M. Malek, K. Martens, 
 C. Mauger, C. McGrew, E. Sharkey, B. Viren, C. Yanagisawa, T. Toshito, C. Mitsuda, K. Miyano,
 C. Saji, T. Shibata, Y. Kajiyama, Y. Nagashima, K. Nitta, M. Takita, H.I. Kim, S.B. Kim,
 J. Yoo, H. Okazawa, T. Ishizuka, M. Etoh, Y. Gando, T. Hasegawa, K. Inoue, K. Ishihara,
 J. Shirai, A. Suzuki, M. Koshiba, Y. Hatakeyama, Y. Ichikawa, M. Koike, K. Nishijima,
 H. Ishino, M. Morii, R. Nishimura, Y. Watanabe, D. Kielczewska,  H.G. Berns,
 S.C. Boyd, A.L. Stachyra, and R.J. Wilkes 
(Super Kamiokande Collaboration): Phys. Lett. B539 (2002) 179. 
\bibitem{SNO1}
Q. R. Ahmad, R. C. Allen, T. C. Andersen, J. D. Anglin, G. Bu"hler, J. C. Barton,
 E. W. Beier, M. Bercovitch, J. Bigu, S. Biller, R. A. Black, I. Blevis, R. J. Boardman,
 J. Boger, E. Bonvin, M. G. Boulay, M. G. Bowler, T. J. Bowles, S. J. Brice,
 M. C. Browne, T. V. Bullard, T. H. Burritt, K. Cameron, J. Cameron, Y. D. Chan, M. Chen,
 H. H. Chen, X. Chen, M. C. Chon, B. T. Cleveland, E. T. H. Clifford, J. H. M. Cowan,
 D. F. Cowen, G. A. Cox, Y. Dai, X. Dai, F. Dalnoki-Veress, W. F. Davidson, P. J. Doe,
 G. Doucas, M. R. Dragowsky, C. A. Duba, F. A. Duncan, J. Dunmore, E. D. Earle,
 S. R. Elliott, H. C. Evans, G. T. Ewan, J. Farine, H. Fergani, A. P. Ferraris, R. J. Ford,
 M. M. Fowler, K. Frame, E. D. Frank, W. Frati, J. V. Germani, S. Gil, A. Goldschmidt,
 D. R. Grant, R. L. Hahn, A. L. Hallin, E. D. Hallman, A. Hamer, A. A. Hamian, R. U. Haq,
 C. K. Hargrove, P. J. Harvey, R. Hazama, R. Heaton, K. M. Heeger, W. J. Heintzelman,
 J. Heise, R. L. Helmer, J. D. Hepburn, H. Heron, J. Hewett, A. Hime, M. Howe,
 J. G. Hykawy, M. C. P. Isaac, P. Jagam, N. A. Jelley, C. Jillings, G. Jonkmans,
 J. Karn, P. T. Keener, K. Kirch, J. R. Klein, A. B. Knox, R. J. Komar, R. Kouzes,
 T. Kutter, C. C. M. Kyba, J. Law, I. T. Lawson, M. Lay, H. W. Lee, K. T. Lesko,
 J. R. Leslie, I. Levine, W. Locke, M. M. Lowry, S. Luoma, J. Lyon, S. Majerus,
 H. B. Mak, A. D. Marino, N. McCauley, A. B. McDonald, D. S. McDonald, K. McFarlane,
 G. McGregor, W. McLatchie, R. Meijer Drees, H. Mes, C. Mifflin, G. G. Miller, G. Milton,
 B. A. Moffat, M. Moorhead, C. W. Nally, M. S. Neubauer, F. M. Newcomer, H. S. Ng,
 A. J. Noble, E. B. Norman, V. M. Novikov, M. O'Neill, C. E. Okada, R. W. Ollerhead,
 M. Omori, J. L. Orrell, S. M. Oser, A. W. P. Poon, T. J. Radcliffe, A. Roberge,
 B. C. Robertson, R. G. H. Robertson, J. K. Rowley, V. L. Rusu, E. Saettler,
 K. K. Schaffer, A. Schuelke, M. H. Schwendener, H. Seifert, M. Shatkay, J. J. Simpson,
 D. Sinclair, P. Skensved, A. R. Smith, M. W. E. Smith, N. Starinsky, T. D. Steiger,
 R. G. Stokstad, R. S. Storey, B. Sur, R. Tafirout, N. Tagg, N. W. Tanner, R. K. Taplin,
 M. Thorman, P. Thornewell, P. T. Trent, Y. I. Tserkovnyak, R. Van Berg, R. G. Van de Water,
 C. J. Virtue, C. E. Waltham, J.-X. Wang, D. L. Wark, N. West, J. B. Wilhelmy,
 J. F. Wilkerson, J. Wilson, P. Wittich, J. M. Wouters, and M. Yeh (SNO Collaboration):
 Phys. Rev. Lett. 87 (2001) 071301.
\bibitem{SNO2}
Q. R. Ahmad, R. C. Allen, T. C. Andersen, J. D.Anglin, J. C. Barton,   E. W. Beier,
 M. Bercovitch, J. Bigu, S. D. Biller, R. A. Black, I. Blevis, R. J. Boardman, J. Boger,
 E. Bonvin, M. G. Boulay,  M. G. Bowler, T. J. Bowles, S. J. Brice,  M. C. Browne,
  T. V. Bullard, G. Bu"hler, J. Cameron, Y. D. Chan, H. H. Chen, M. Chen, X. Chen, 
 B. T. Cleveland, E. T. H. Clifford, J. H. M. Cowan, D. F. Cowen, G. A. Cox, X. Dai,
 F. Dalnoki-Veress, W. F. Davidson, P. J. Doe,   G. Doucas, M. R. Dragowsky,
  C. A. Duba, F. A. Duncan, M. Dunford, J. A. Dunmore, E. D. Earle,  S. R. Elliott,
  H. C. Evans, G. T. Ewan, J. Farine,  H. Fergani, A. P. Ferraris, R. J. Ford,
 J. A. Formaggio, M. M. Fowler, K. Frame, E. D. Frank, W. Frati, N. Gagnon,
    J. V. Germani, S. Gil, K. Graham, D. R. Grant, R. L. Hahn, A. L. Hallin,
 E. D. Hallman, A. S. Hamer,  A. A. Hamian, W. B. Handler, R. U. Haq, C. K. Hargrove,
 P. J. Harvey, R. Hazama, K. M. Heeger, W. J. Heintzelman, J. Heise,  R. L. Helmer,
  J. D. Hepburn, H. Heron, J. Hewett, A. Hime, M. Howe, J. G. Hykawy, M. C. P. Isaac,
 P. Jagam, N. A. Jelley, C. Jillings, G. Jonkmans, K. Kazkaz, P. T. Keener,
 J. R. Klein, A. B. Knox, R. J. Komar, R. Kouzes, T. Kutter, C. C. M. Kyba, J. Law,
 I. T. Lawson, M. Lay, H. W. Lee, K. T. Lesko, J. R. Leslie, I. Levine, W. Locke,
 S. Luoma, J. Lyon, S. Majerus, H. B. Mak, J. Maneira, J. Manor, A. D. Marino,
 N. McCauley,  A. B. McDonald,  D. S. McDonald, K. McFarlane, G. McGregor,
 R. Meijer Drees, C. Mifflin, G. G. Miller, G. Milton, B. A. Moffat, M. Moorhead,
 C. W. Nally, M. S. Neubauer, F. M. Newcomer, H. S. Ng, A. J. Noble, E. B. Norman,
 V. M. Novikov, M. O'Neill, C. E. Okada, R. W. Ollerhead, M. Omori, J. L. Orrell,
 S. M. Oser, A. W. P. Poon,    T. J. Radcliffe, A. Roberge, B. C. Robertson,
 R. G. H. Robertson,  S. S. E. Rosendahl, J. K. Rowley, V. L. Rusu, E. Saettler,
 K. K. Schaffer, M. H. Schwendener, A. Schu"lke, H. Seifert,   M. Shatkay, J. J. Simpson,
 C. J. Sims, D. Sinclair,  P. Skensved, A. R. Smith, M. W. E. Smith, T. Spreitzer,
 N. Starinsky, T. D. Steiger, R. G. Stokstad, L. C. Stonehill, R. S. Storey, B. Sur,
  R. Tafirout, N. Tagg,  N. W. Tanner, R. K. Taplin, M. Thorman, P. M. Thornewell,
 P. T. Trent, Y. I. Tserkovnyak, R. Van Berg, R. G. Van de Water,  C. J. Virtue,
 C. E. Waltham, J.-X. Wang, D. L. Wark, N. West, J. B. Wilhelmy, J. F. Wilkerson,
  J. R. Wilson, P. Wittich ,   J. M. Wouters, and M. Yeh   (SNO Collaboration):
 Phys. Rev. Lett. 89 (2002) 011301. 
\bibitem{KamLAND1}
K. Eguchi, S. Enomoto, K. Furuno, J. Goldman, H. Hanada, H. Ikeda, K. Ikeda, K. Inoue, 
K. Ishihara, W. Itoh, T. Iwamoto, T. Kawaguchi, T. Kawashima, H. Kinoshita, Y. Kishimoto,
 M. Koga, Y. Koseki, T. Maeda, T. Mitsui, M. Motoki, K. Nakajima, M. Nakajima, T. Nakajima,
 H. Ogawa, K. Owada, T. Sakabe, I. Shimizu, J. Shirai, F. Suekane, A. Suzuki, K. Tada,
 O. Tajima, T. Takayama, K. Tamae, H. Watanabe, J. Busenitz, Z. Djurcic, K. McKinny,
 D.-M. Mei, A. Piepke, E. Yakushev,    E. Berger, Y. D. Chan, M. P. Decowski, D. A. Dwyer,
 S. J. Freedman, Y. Fu, B. K. Fujikawa, K. M. Heeger, K. T. Lesko, K.-B. Luk, H. Murayama,
 D. R. Nygren, C. E. Okada, A. W. P. Poon, H. M. Steiner, L. A. Winslow,
 G. A. Horton-Smith, R. D. McKeown, J. Ritter, B. Tipton, P. Vogel, C. E. Lane, T. Miletic,
    W. Gorham, G. Guillian, J. G. Learned, J. Maricic, S. Matsuno, S. Pakvasa, S. Dazeley,
 S. Hatakeyama, M. Murakami, R. C. Svoboda, B. D. Dieterle, M. DiMauro, J. Detwiler,
 G. Gratta, K. Ishii, N. Tolich, Y. Uchida, M. Batygov, W. Bugg, H. Cohn, Y. Efremenko,
 Y. Kamyshkov, A. Kozlov, Y. Nakamura, L. De Braeckeleer, C. R. Gould, H. J. Karwowski,
 D. M. Markoff, J. A. Messimore, K. Nakamura, R. M. Rohm, W. Tornow, A. R. Young,
 and Y.-F. Wang    (KamLAND Collaboration): Phys. Rev. Lett. 90 (2003) 021802.
\bibitem{KamLAND2}
T. Araki, K. Eguchi, S. Enomoto, K. Furuno, K. Ichimura, H. Ikeda, K. Inoue, K. Ishihara,
   T. Iwamoto,    T. Kawashima, Y. Kishimoto, M. Koga, Y. Koseki, T. Maeda, T. Mitsui,
 M. Motoki, K. Nakajima, H. Ogawa, K. Owada, J.-S. Ricol, I. Shimizu, J. Shirai, F. Suekane,
 A. Suzuki, K. Tada, O. Tajima, K. Tamae, Y. Tsuda, H. Watanabe, J. Busenitz, T. Classen,
 Z. Djurcic, G. Keefer, K. McKinny, D.-M. Mei,    A. Piepke, E. Yakushev, B. E. Berger,
 Y. D. Chan, M. P. Decowski, D. A. Dwyer, S. J. Freedman, Y. Fu, B. K. Fujikawa, J. Goldman,
 F. Gray, K. M. Heeger, K. T. Lesko, K.-B. Luk, H. Murayama,    A. W. P. Poon,
 H. M. Steiner, L. A. Winslow, G. A. Horton-Smith,    C. Mauger, R. D. McKeown, P. Vogel,
 C. E. Lane, T. Miletic, P. W. Gorham, G. Guillian, J. G. Learned, J. Maricic, S. Matsuno,
 S. Pakvasa, S. Dazeley, S. Hatakeyama, A. Rojas, R. Svoboda, B. D. Dieterle, J. Detwiler,
  G. Gratta,  K. Ishii, N. Tolich, Y. Uchida,      M. Batygov, W. Bugg, Y. Efremenko,
 Y. Kamyshkov, A. Kozlov, Y. Nakamura, C. R. Gould, H. J. Karwowski, D. M. Markoff,
 J. A. Messimore, K. Nakamura, R. M. Rohm, W. Tornow, R. Wendell, A. R. Young, M.-J. Chen,
 Y.-F. Wang, and F. Piquemal    (KamLAND Collaboration): Phys. Rev. Lett. 94 (2005) 081801. 
\bibitem{pdg}
K. Nakamura, K. Hagiwara,   K. Hikasa,  H. Murayama,  M. Tanabashi,  T. Watari, C. Amsler, 
  M. Antonelli,     D.M. Asner,    H. Baer,      H.R. Band,      R.M. Barnett,
  T. Basaglia,      E. Bergren,      J. Beringer,      G. Bernardi,      W. Bertl, 
  H. Bichsel,      O. Biebel,      E. Blucher,      S. Blusk,      R.N. Cahn,     M. Carena,
 A. Ceccucci,    D. Chakraborty,    M.-C. Chen,    R.S. Chivukula,   G. Cowan,  O. Dahl,   
 G. D'Ambrosio,     T. Damour,    D. de Florian,   A. de Gouvea,    T. DeGrand,  
  G. Dissertori,   B. Dobrescu,   M. Doser,   M. Drees,    D.A. Edwards,    S. Eidelman,
 J. Erler,   V.V. Ezhela,    W. Fetscher,   B.D. Fields,   B. Foster,    T.K. Gaisser,
  L. Garren,   H.-J. Gerber,  G. Gerbier,   T. Gherghetta,   G.F. Giudice,  S. Golwala,
  M. Goodman, C. Grab,     A.V. Gritsan,   J.-F. Grivaz,  D.E. Groom,   M. Gru"newald,
   A. Gurtu,   T. Gutsche,  H.E. Haber,  C. Hagmann,    K.G. Hayes,   M. Heffner,  B. Heltsley,
  J.J. Herna'ndez-Rey,    A. Ho"cker,   J. Holder,   J. Huston,  J.D. Jackson,   K.F. Johnson,
  T. Junk,   A. Karle,  D. Karlen,  B. Kayser,   D. Kirkby,  S.R. Klein,  C. Kolda,  
 R.V. Kowalewski,  B. Krusche,   Yu.V. Kuyanov,  Y. Kwon,   O. Lahav,   P. Langacker,  A. Liddle,   
 Z. Ligeti,  C.-J. Lin,  T.M. Liss,    L. Littenberg,   K.S. Lugovsky,  S.B. Lugovsky,   J. Lys, 
 H. Mahlke,  T. Mannel,  A.V. Manohar,  W.J. Marciano,   A.D. Martin,  A. Masoni,  D. Milstead,  
  R. Miquel,  K. Mo"nig,   M. Narain,  P. Nason,  S. Navas,   P. Nevski,  Y. Nir,  K.A. Olive,
  L. Pape,  C. Patrignani,  J.A. Peacock,  S.T. Petcov,    A. Piepke,  G. Punzi,  A. Quadt,
 S. Raby,   G. Raffelt,  B.N. Ratcliff,  P. Richardson,   S. Roesler,  S. Rolli,  A. Romaniouk, 
  L.J. Rosenberg,   J.L. Rosner,  C.T. Sachrajda,  Y. Sakai,  G.P. Salam, S. Sarkar,  
 F. Sauli,   O. Schneider,   K. Scholberg,  D. Scott,   W.G. Seligman,  M.H. Shaevitz,
  M. Silari,  T. Sjo"strand,  J.G. Smith, G.F. Smoot,  S. Spanier,  H. Spieler,   A. Stahl,
  T. Stanev,    S.L. Stone,   T. Sumiyoshi,   M.J. Syphers,   J. Terning,    M. Titov,  
 N.P. Tkachenko,  N.A. To"rnqvist, D. Tovey,   T.G. Trippe,  G. Valencia,   K. van Bibber,
  G. Venanzoni,   M.G. Vincter,   P. Vogel,   A. Vogt,   W. Walkowiak,   C.W. Walter,
 D.R. Ward,  B.R. Webber,  G. Weiglein,   E.J. Weinberg,  J.D. Wells,    A. Wheeler,  
  L.R. Wiencke,  C.G. Wohl,   L. Wolfenstein,   J. Womersley,    C.L. Woody,   R.L. Workman, 
  A. Yamamoto,  W. -M. Yao,  O.V. Zenin,  J. Zhang,     R.-Y. Zhu  and   P.A. Zyla 
      (Particle Data Group): J. Phys. G 37 (2010) 075021.

\bibitem{nulifetime1}
M. T. Ressell and M. S. Turner: Comm. Astrophys. 14 (1990) 323.
\bibitem{nulifetime2}
A. Mirizzi, D. Montanino and P. D. Serpico: Phys. Rev. D76 (2007) 053007.
 
\bibitem{COBE1}
M. G. Hauser, R. G. Arendt, T. Kelsall, E. Dwek, N. Odegard, J. L. Weiland, H. T. Freudenreich,
W. T. Reach, R. F. Silverberg, S. H. Moseley, Y. C. Pei, P. Lubin, J. C. Mather, R. A. Shafer,
G. F. Smoot, R. Weiss, D. T. Wilkinson and E. L. Wright: Astrophys. J. 508 (1998) 25.
\bibitem{COBE2}
D. P. Finkbeiner, M. Davis and D. J. Schlegel: Astrophys. J. 544 (2000) 81.   

\bibitem{AKARI} 
S. Matsuura, M. Shirahata, M. Kawada, T. T. Takeuchi, D. Burgarella, D. L. Clements, 
W. S. Jeong, H. Hanami, S. A. Khan, H. Matsuhara, T. Nakagawa, S. Oyabu, C. P. Pearson,
A. Pollo, S. Serjeant, T. Takagi and G. White: Astrophys. J. 737 (2011) 2.

\bibitem{spitzer} 
H. Dole, G. Lagache, J. L. Puget, K. I. Caputi, N. Fernandez-Conde, E. Le Floc'h, C. Papovich,
P. G. Perez-Gonzalez, G. H. Rieke and M. Blaylock: Astron. Astrophys. 451 (2006) 417-429.

\bibitem{hershel}
 S. Berta, B. Magnelli, D. Lutz, B. Altieri, H. Aussel, P. Andreani, O. Bauer, A. Bongiovanni,
 A. Cava, J. Cepa, A. Cimatti, E. Daddi, H. Dominguez, D. Elbaz, H. Feuchtgruber,
 N. M. Fo"rster Schreiber, R. Genzel, C. Gruppioni, R. Katterloher, G. Magdis, R. Maiolino,
 R. Nordon, A. M. Pe'rez Garci'a, A. Poglitsch, P. Popesso, F. Pozzi, L. Riguccini,
 G. Rodighiero, A. Saintonge, P. Santini, M. Sanchez-portal, L. Shao, E. Sturm, L. J. Tacconi,
 I. Valtchanov, M. Wetzstein and E. Wieprecht: Astron. Astrophys. 518 (2010) L30.

\bibitem{SM_nudecay1}
P. B. Pal and L. Wolfenstein:  Phys. Rev. D25 (1982) 766.
\bibitem{SM_nudecay2}
K. Sato and M. Kobayashi: Prog. Theor. Phys. 58 (1977) 1775. 

\bibitem{LR_model1}
R. E. Schrock: Nucl. Phys. B206 (1982) 359.
\bibitem{LR_model2}
M. A. B. Beg and W. J. Marciano: Phys. Rev. D17 (1978) 1395.

\bibitem{MNS}
Z. Maki, M. Nakagawa and S. Sakata:  Prog. Theor. Phys. 28 (1962) 870.

\bibitem{spica} T. Nakagawa:
American Astronomical Society, AAS Meeting \# 218, \# 314.01; Bulletin of the American 
Astronomical Society, Vol. 43, 2011.

\bibitem{exzit}S. Matsuura:
Proceedings of Far-IR, Sub-mm and MM Detector Technology Workshop, NASA/CP-211408, 2002. 

\end{thebibliography}
\end{document}